\documentclass[10pt,aps,prl,twocolumn,showpacs]{revtex4}

\usepackage{graphicx}
\begin{document}

\title{Minimizing the linewidth of the Flux-Flow Oscillator}
\author{A.L. Pankratov}
\email{alp@ipm.sci-nnov.ru}
\affiliation{Institute for Physics of Microstructures of RAS, Nizhny Novgorod, Russia}

\begin{abstract}
For the first time the linewidth of Flux-Flow Oscillator has been
calculated by direct computer simulation of the sine-Gordon equation
with noise. Nearly perfect agreement of the numerical results with
the formula derived in [Phys. Rev. B, {\bf 65}, 054504 (2002)] has
been achieved.  It has been demonstrated that for homogeneous bias
current distribution the linewidth actually does not depend on the
junction length for practically interesting parameters range.
Depending on the length of the unbiased tail, the power may be
maximized and the linewidth may be minimized in a broad range of
bias currents. The linewidth can be decreased further by 1.5 times
by proper load matching.
\end{abstract}

%\begin{keyword}
%Tunneling, Josephson effect, Josephson devices

% PACS codes here, in the form:
%PACS $74.50.+r$ $03.75.Lm$
%\end{keyword}
\maketitle

During the last decade the flux-flow oscillator (FFO), based on a
viscous flow of magnetic flux quanta in a long Josephson tunnel
junction (JTJ) \cite{naga}, has been considered as the most
promising local oscillator in superconducting spectrometers
\cite{KS} for space-born radio astronomy and atmosphere monitoring
due to its wide operational bandwidth and easy broadband tunability.
However, the spectral linewidth of the emitted radiation of the
free-running FFO is rather large, that complicates the phase
locking. Typically, the free-running linewidth is 2-20 MHz for an
Nb-AlOx-Nb FFO in the 400-700GHz frequency range. For spectral
applications it is of crucial importance to reduce the FFO
linewidth, to make it more homogeneous in all working frequency
range, and to increase the emitted power to improve the
signal-to-noise ratio.

The dynamical and fluctuational properties of the FFO have been
investigated in \cite{naga}-\cite{Fed}, in particular, the linewidth
has been studied both experimentally \cite{kosh1}-\cite{KS052} and
theoretically \cite{gmu96}-\cite{myg05}. However, the only formula
for the linewidth, derived in \cite{alplw}, that takes into account
not only differential resistance over bias current, but also
differential resistance over magnetic field, has been proven as the
most adequately describing experimental results, see
\cite{KSsust},\cite{PhysC02}. But, the nature of conversion of bias
current fluctuations to the magnetic field fluctuations is still
unclear, and the conversion factor is not known exactly. Also, the
dependence of the linewidth on the bias current profile and certain
parameters, such as RC-load, has not been systematically studied yet
neither theoretically nor experimentally. The aim of the present
paper is to study the FFO linewidth by direct computer simulation of
the sine-Gordon equation with noise, to compare the obtained results
with the formula of \cite{alplw} and to make certain optimizations
of FFO design (varying bias current profile and the RC-load) in
order to minimize the linewidth and to increase the emitted power.

For several decades the sine-Gordon model has been the most adequate
model for the long Josephson junction, giving a good qualitative
description of its basic properties:
\begin{equation}
{\phi}_{tt}+\alpha{\phi}_{t}
-{\phi}_{xx}=\beta{\phi}_{xxt}+\eta(x)-\sin (\phi)+\eta_f(x,t),
\label{PSGE}
\end{equation}
where indices $t$ and $x$ denote temporal and spatial derivatives,
respectively. Space and time are normalized to the Josephson
penetration length $\lambda _{J}$ and to the inverse plasma
frequency $\omega_{p}^{-1}$, respectively,
$\alpha={\omega_{p}}/{\omega_{c}}$ is the damping parameter, where
$\omega_p=\sqrt{2eI_c/\hbar C}$, $\omega_{c}=2eI_cR_{N}/\hbar$,
$I_c$ is the critical current, $C$ is the JTJ capacitance, and $R_N$
is the normal state resistance, $\beta$ is the surface loss
parameter, $\eta(x)$ is the dc overlap bias current density,
normalized to the critical current density $J_c$, and $\eta_f(x,t)$
is the fluctuational current density. In the case where the
fluctuations are treated as white Gaussian noise with zero mean, and
the critical current density is fixed, its correlation function is:
$\left<i_f(x,t)i_f(x',t')\right>=2\alpha\gamma \delta
(x-x^{\prime})\delta (t-t^{\prime})$, where $\gamma = I_{T} /
(J_{c}\lambda_J)$ is the dimensionless noise intensity \cite{Fed},
$I_{T}=2ekT/\hbar$ is the thermal current, $e$ is the electron
charge, $\hbar$ is the Planck constant, $k$ is the Boltzmann
constant and $T$ is the temperature.

The boundary conditions, that simulate simple RC-loads, see Ref.s
\cite{Parment} and \cite{pskm}, have the form:
\begin{eqnarray}\label{x=0}
\phi(0,t)_{x}+r_L c_L\phi(0,t)_{xt}-c_L\phi(0,t)_{t t}+\\
\beta r_R c_R\phi(0,t)_{xtt}+\beta\phi(0,t)_{x
t}=\Gamma-\Delta\Gamma, \nonumber \\ \phi(L,t)_{x}+r_R
c_R\phi(L,t)_{x t}+c_R\phi(L,t)_{t t}+ \label{x=L} \\  \beta r_R
c_R\phi(L,t)_{xtt}+\beta\phi(L,t)_{x t}=\Gamma+\Delta\Gamma.
\nonumber
\end{eqnarray}
Here $\Gamma$ is the normalized magnetic field,
$\Delta\Gamma=0.05\Gamma$ is a small magnetic field difference, see
\cite{pskm}, and $L$ is the dimensionless length of JTJ. The
dimensionless capacitances and resistances, $c_{L,R}$ and $r_{L,R}$,
are the FFO RC-load placed at the left (output) and at the right
(input) ends, respectively. It should be noted that, following Ref.
\cite{myg05}, if both overlap $\eta_{ov}=(1/L)\int_0^L\eta(x)dx$ and
inline $\eta_{in}=2\Delta\Gamma/L$ components of the current are
present, the total current, $\eta_t$, with respect to which all
current-voltage characteristics will be computed, is the sum of
overlap and inline components: $\eta_t=\eta_{ov}+\eta_{in}$.

\begin{figure}[h]
\resizebox{1\columnwidth}{!}{
\includegraphics{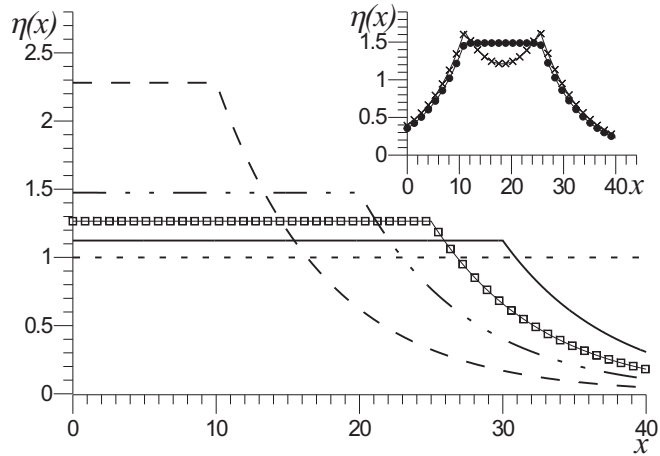}}
{\caption{ The distribution of overlap component of bias current
$\eta(x)$. Short-dashed line - homogeneous distribution; solid curve
- $x_1=30$; curve with rectangles - $x_1=25$; dot-dashed curve -
$x_1=20$; long-dashed curve - $x_1=10$. Inset: curve with circles -
$x_0=11$, $x_1=25.5$, $a=0$; curve with crosses - $x_0=11$,
$x_1=25.5$, $a=0.005$.} \label{Fig1}}
\end{figure}

In Ref. \cite{pskm} on the basis of Eq. (\ref{PSGE}) without noise
and boundary conditions (\ref{x=0}), (\ref{x=L}), the investigation
of current-voltage characteristics of FFO has been performed. For
the bias current profile, depicted in the inset of Fig. \ref{Fig1}
by curve with crosses, good qualitative agreement with experimental
IVCs has been achieved. Due to experimental motivation it was
assumed, that the current profile was parabolic (with the curvature
$a=0.005$) between the left and the right boundaries of bias
electrode $x_0$ and $x_1$ ($0\le x_0\le x_1 \le L$), and drops down
exponentially in the unbiased tails $x\le x_0$, $x\ge x_1$ with the
decay factor $p$: $\exp(-px)$ (with $p=0.13$ in Fig. 1). The decay
factor and the parabolic curvature were used as fitting parameters
when the comparison with the experimental IVCs was done.

The key question arises: if one will change the lengths of unbiased
tails, how it will affect the emitted power and the linewidth? In
Ref. \cite{naga} it was suggested to use the unbiased tail to
decrease the differential resistance $r_d$ that might reduce the
linewidth, if the formula for the linewidth of short JTJ would work
for FFO (here and below the linewidth is defined as full width, half
power):
\begin{equation}\label{slw}
\Delta f_s=2\alpha\gamma r_d^2/L.
\end{equation}
Later, it was found experimentally \cite{UstinovSP}, that even for
small $r_d$ the FFO linewidth is almost one order of magnitude
larger than predicted by (\ref{slw}). In Ref. \cite{alplw}, the
formula for the FFO linewidth, that takes into account not only
conventional differential resistance over bias current, but also
differential resistance over magnetic field $r_d^{CL}=Ldv/d\Gamma$
(control line current) was derived:
\begin{equation}
\Delta f_{FFO}=2\alpha\gamma (r_d+\sigma r^{CL}_d)^2 /L,
\label{lwffo}
\end{equation}
and demonstrated good agreement with experiment
\cite{KSsust},\cite{PhysC02}. In Ref. \cite{alput} it has been
demonstrated that by the choice of the bias current profile, the
radiation can be either enhanced or suppressed, and it is desirable
to supply more bias current at the radiating end, than at the input
end. Therefore, in the frame of the present paper we shift the
current profile to the left, $x_0=0$, and vary the length of
unbiased tail, which is located at the right end of JTJ. Also, to
avoid problems with scaling of parabolic curvature (it is not clear
what to keep constant, the curvature or the depth of the "well"),
let us set the current profile to be constant between $x_0$ and
$x_1$.

\begin{figure}[h]
\resizebox{1\columnwidth}{!}{
\includegraphics{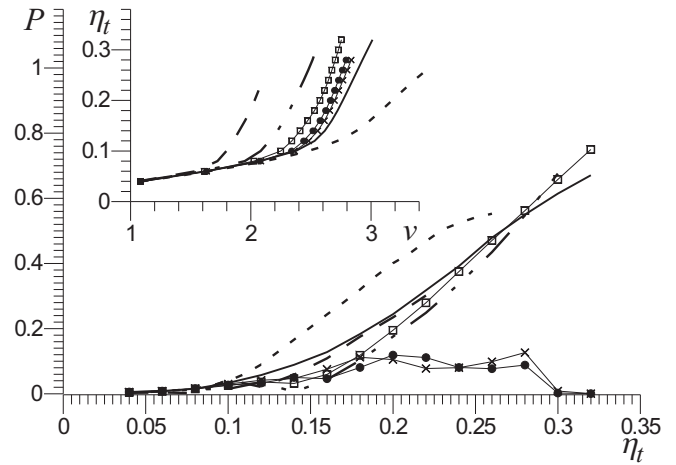}}
{\caption{Radiated power versus total current, computed for
$\eta(x)$, presented in Fig. \ref{Fig1}: short-dashed line -
homogeneous distribution; solid curve - $x_1=30$; curve with
rectangles - $x_1=25$; dot-dashed curve - $x_1=20$; long-dashed
curve - $x_1=10$; curve with circles - $x_0=11$, $x_1=25.5$, $a=0$;
curve with crosses - $x_0=11$, $x_1=25.5$, $a=0.005$. Inset: dc
current-voltage characteristics computed for $\eta(x)$, presented in
Fig. \ref{Fig1}: the notations are the same as for power.}
\label{FigCV}}
\end{figure}

The power at RC-load at the radiating end $x=0$ for different bias
current profiles depicted in Fig. \ref{Fig1}, is presented in
Fig.\ref{FigCV}. The power is computed in accordance with
\cite{Parment}. The implicit difference scheme, used to solve Eq.
(\ref{PSGE}) with noise, has been successfully tested in \cite{Fed}
when the mean escape times from zero voltage state were
investigated. The parameters are the following: $L=40$,
$\alpha=0.033$, $\beta=0.035$, $c_L=c_R=100$, $r_L=2$, $r_R=100$,
$\Gamma=3.6$, and $\gamma=0.1$. From Fig. \ref{FigCV} one can see,
that for the case of two unbiased tails $x_0=11$, $x_1=25.5$ the
power is minimal and almost order of magnitude smaller than for the
current profile with the one unbiased tail $x_1=25$, which gives
maximal power among all considered current profiles. In the inset of
Fig. \ref{FigCV} the current-voltage characteristics for the same
current profiles are given for comparison: it is seen, that the
flux-flow steps have largest height also for $x_1=25$. The height of
IVCs for both profiles with $x_0=11$, $x_1=25.5$ and $a=0.005$,
$a=0$ have close values to each other and are comparable to
$x_1=25$. So, it is desirable to give larger bias current at the
radiating end to get larger emitted power. It should be noted that
for both profiles with two unbiased tails, the power versus bias
current has minima as in experiment \cite{KS052}. However, the
investigation of this phenomenon is out of scope of the present
paper and will be presented elsewhere.
\begin{figure}[h]
\resizebox{1\columnwidth}{!}{
\includegraphics{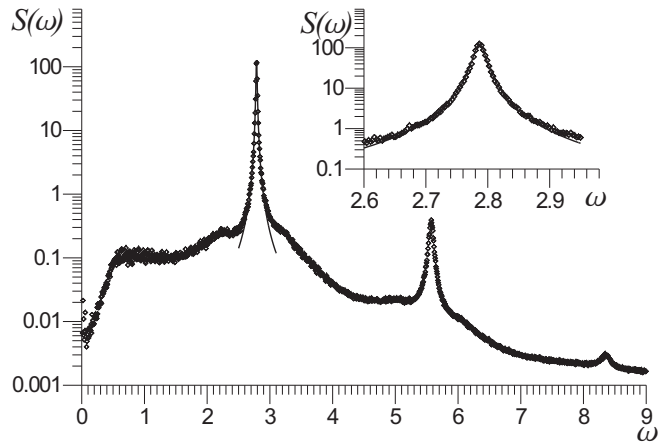}}
{\caption{The calculated spectral density for $x_1=25$,
$\eta_0=0.3$, $\gamma=0.1$ - diamonds; Lorentzian approximation -
solid curve. Inset: the same, enlarged around the spike.}
\label{FigSP}}
\end{figure}

The power spectral density of FFO was computed as Fourier transform
of the correlation function of the second kind
$\Phi[\tau]=\frac{1}{T_{av}}\int_0^{T_{av}}\left<v_0(t)v_0(t+\tau)\right>dt$,
where $v_0(t)=d\varphi(t,0)/dt$ is the voltage at the RC-load
($x=0$) and $T_{av}$ is the averaging time. There are two general
restrictions, that complicates the calculation of the spectral
density and the linewidth: on one hand, the time step should be
small enough to resolve oscillations, and the averaging time
$T_{av}$ should be rather large to resolve fine spectral spikes. Due
to these restrictions the noise intensity was chosen $\gamma=0.1$.
Nevertheless, this is the same limit of low noise intensity as in
experiments, since IVCs are almost unaffected by noise, the spectral
spikes are narrow, and the linewidth perfectly scales proportionally
to the noise intensity, see below. For the linewidth calculations
the following parameters were used: spatial step $\Delta x=0.05$,
temporal step $\Delta t\approx 0.1$, $T_{av}=8000$.
\begin{figure}[h]
\resizebox{1\columnwidth}{!}{
\includegraphics{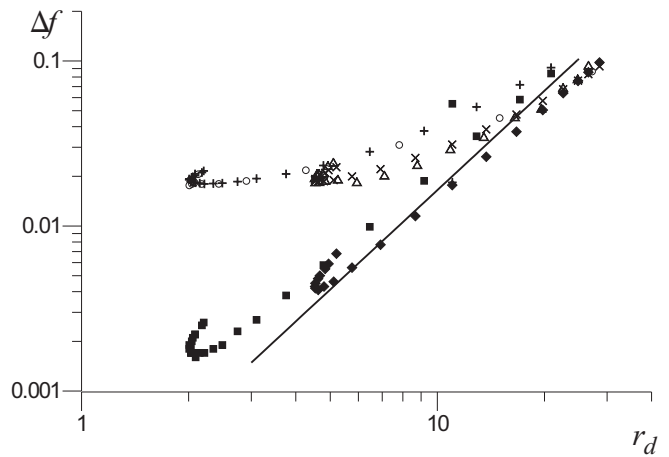}}
{\caption{FFO linewidth versus differential resistance $r_d$ for
$L=40$ and $\gamma=0.1$. Empty triangles and crosses - simulations
and theory (\ref{lwffo}) for homogeneous bias current distribution,
empty circles and crests - simulations and theory for the case with
unbiased tail $x_1=30$, filled diamonds and rectangles - theory of
\cite{ssy}, solid line - theory (\ref{slw}).} \label{FigLWrd}}
\end{figure}

In Fig. \ref{FigSP} the power spectral density of FFO is presented.
As one can see, the emitted signal is nearly sinusoidal, in
agreement with \cite{Parment} and experimental results: the power
contained in the second and third harmonics is much lower than in
the main one. Also, the spectral peak is perfectly Lorentzian in
more than two orders of magnitude interval.
\begin{figure}[ht]
\resizebox{1\columnwidth}{!}{
\includegraphics{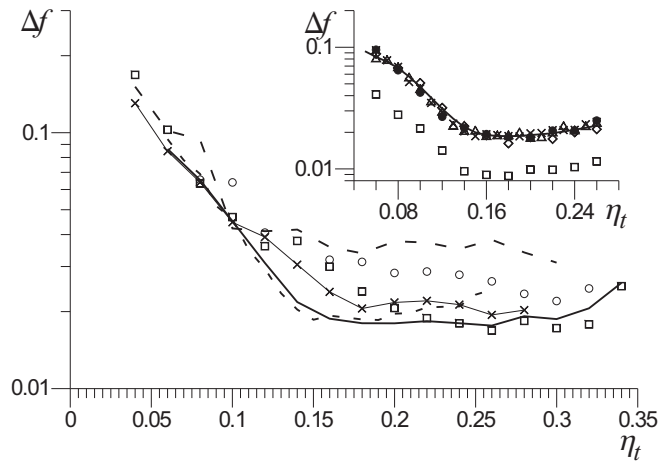}}
{\caption{FFO linewidth versus total current, computed for
$\eta(x)$, presented in Fig. \ref{Fig1}, $\gamma=0.1$: short-dashed
curve - homogeneous distribution; solid curve - $x_1=30$; rectangles
- $x_1=25$; circles - $x_1=20$; long-dashed curve - $x_1=10$; curve
with crosses - $x_0=11$, $x_1=25.5$, $a=0.005$. Inset: FFO linewidth
for junctions with different lengths: $L=20$ - diamonds, $L=40$ -
crosses, $L=60$ - circles, $L=80$ - triangles, solid curve - theory
for $L=60$, all these curves for $\gamma=0.1$; rectangles -
simulations for $L=40$, $\gamma=0.05$. } \label{FigLWI}}
\end{figure}

In Fig. \ref{FigLWrd} the FFO linewidth versus differential
resistance $r_d$ for junctions of $L=40$ and $\gamma=0.1$ is
presented for the case of homogeneous bias current distribution and
$x_1=30$. It is seen, that in both cases formula (\ref{lwffo}) is in
good agreement with numerical results, while formula (\ref{slw}) and
formula from \cite{ssy} (in formula (\ref{slw}) $\alpha$ must be
substituted by inverse static resistance $\eta_t/v$) significantly
underestimate the linewidth.

The appearance of the conversion of bias current fluctuations to
magnetic field fluctuations may be explained in the following way:
boundary conditions (\ref{x=0}), (\ref{x=L}) of Eq. (\ref{PSGE})
depend on the phase, which fluctuates, since it is governed by Eq.
(\ref{PSGE}) with noise. It is quite difficult to calculate the
noise conversion factor $\sigma$, since it should be done
self-consistently. In the following, to see how the linewidth
behaves in all working range, let us plot it versus total bias
current $\eta_t$.

Let us analyze the dependence of the linewidth on the length of
unbiased tail. From Fig. \ref{FigLWI} it is seen, that minimal value
of $\Delta f$ is reached for several cases, including the
homogeneous one. However, for the unbiased tail $x_1=30$, the
linewidth is nearly constant (and minimal) in a maximal range of
bias currents. Therefore, the unbiased tail of 1/4 of junction's
length gives the nearly maximal power and nearly minimal linewidth
in broadest range of bias current, and can be recommended for
spectral applications.

Let us investigate the dependence of the linewidth versus junction's
length $L$ for homogeneous bias current distribution. The
corresponding curves are presented in the inset of Fig.
\ref{FigLWI}. It is seen, that increase of $L$ does not help to
decrease the linewidth, the curves nearly coincide. The lowest curve
is computed for noise intensity $\gamma=0.05$ and the linewidth is
two times smaller than for $\gamma=0.1$. Therefore, the noise
intensity $\gamma=0.1$ is indeed rather low, and in the following
one can get good estimate for experimental parameters by
multiplication of the computed curve on the numerical factor
$\gamma_e/\gamma$, where $\gamma_e$ is the noise intensity,
corresponding to the experiment.
\begin{figure}[ht]
\resizebox{1\columnwidth}{!}{
\includegraphics{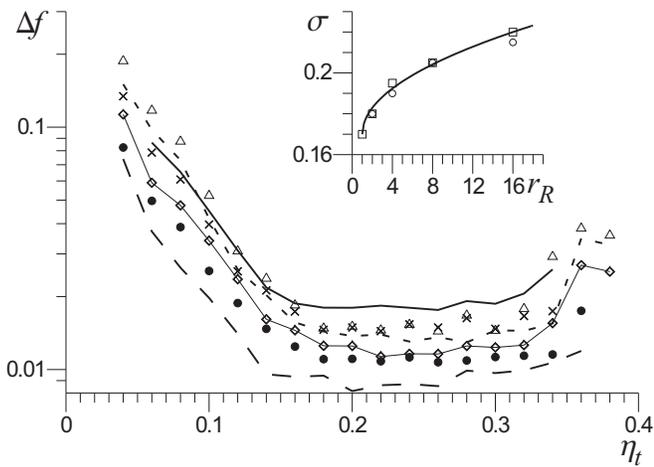}}
{\caption{FFO linewidth versus total current computed for different
values of load resistances and noise intensity, $x_1=30$: solid
curve - $r_L=2$, $r_R=100$; triangles - $r_L=r_R=8$; crosses -
$r_L=2$, $r_R=8$;  short-dashed curve - $r_L=r_R=4$; curve with
diamonds - $r_L=r_R=2$; circles - $r_L=r_R=1$; all these curves for
$\gamma=0.1$; long-dashed curve - $r_L=2$, $r_R=100$ for
$\gamma=0.05$. Inset: the coefficient $\sigma$ versus $r_R$,
rectangles - $r_L=r_R$; circles - $r_L=2$; solid curve - fitting by
$\sqrt{r_R}$.} \label{FigLWIR}}
\end{figure}

Normally, FFO radiating end $x=0$ is well matched to the external
environment, while the opposite end is strongly mismatched, as it
was modeled in \cite{pskm} and in the present paper. It is
interesting to analyze, how the linewidth will change, if the FFO is
better matched at both ends. The results of this analysis are
presented in Fig. \ref{FigLWIR} for $x_1=30$. It is seen, that
improved matching at the opposite end can decrease the linewidth by
1.5 times. It is important to note, that equal matching at both ends
gives smaller linewidth than perfect matching at the radiating end
and bad at the opposite one, e.g., compare the curves for $r_L=2$,
$r_R=100$ and $r_L=r_R=4$ and note, that the curves for $r_L=r_R=8$
and $r_L=2$, $r_R=8$ nearly coincide. Finally, it is important to
mention, that the noise conversion factor $\sigma$ perfectly scales
as $\sqrt{r_R}$ both for $r_L=2$, and for $r_L=r_R$.

The linewidth of Flux-Flow Oscillator has been calculated by direct
computer simulation of the modified sine-Gordon equation with noise
which takes into account surface losses and RC load. Nearly perfect
agreement of the computer simulation results with the formula
(\ref{lwffo}) has been achieved.  It has been demonstrated that for
homogeneous bias current distribution the linewidth actually does
not depend on the junction length for practically interesting
parameters range. Depending on the length of the unbiased tail, the
power may be maximized and the linewidth may be minimized in a broad
range of bias currents. The linewidth can be decreased further by
1.5 times by proper load matching.

The work was supported by the ISTC project 3174.

\end{document}